\documentclass[twocolumn,prl,showpacs,10pt]{revtex4}
\usepackage{graphicx}
\usepackage{epsfig}
\usepackage{amsmath, amssymb}

\newcommand{\lsim}{\,\raise 0.4ex\hbox{$<$}\kern -0.8em\lower 0.62ex\hbox{$\sim$}\,}
\newcommand{\gsim}{\,\raise 0.4ex\hbox{$>$}\kern -0.7em\lower 0.62ex\hbox{$\sim$}\,}

\newcommand{\cs}{c_s^2}

\def\be{\begin{equation}}
\def\ee{\end{equation}}
\def\bea{\begin{eqnarray}}
\def\eea{\end{eqnarray}}
\def\be{\begin{equation}}
\def\be{\begin{equation}}
\newcommand\de{\mathrm{DE}}

\begin{document}

\title{Dark Energy versus Modified Gravity}
\date{December 17, 2006}

\author{Martin Kunz}
\email{Martin.Kunz@physics.unige.ch}
\affiliation{D\'epartement de
Physique Th\'eorique, Universit\'e de
Gen\`eve, 24 quai Ernest Ansermet, CH--1211 Gen\`eve 4, Switzerland}
\author{Domenico Sapone}
\email{Domenico.Sapone@physics.unige.ch}
\affiliation{D\'epartement de
Physique Th\'eorique, Universit\'e de
Gen\`eve, 24 quai Ernest Ansermet, CH--1211 Gen\`eve 4, Switzerland}

\begin{abstract}
There is now strong observational evidence that the expansion of the universe
is accelerating. The standard explanation invokes an unknown ``dark energy''
component. But such scenarios are faced with serious theoretical problems, which has
led to increased interest in models where instead General Relativity is
modified in a way that leads to the observed accelerated expansion. The
question then arises whether the two scenarios can be distinguished. Here
we show that this may not be so easy, demonstrating explicitely that a
generalised dark energy model can match the growth rate of the DGP model
and reproduce the 3+1 dimensional metric perturbations. Cosmological
observations are then unable to distinguish the two cases.
\end{abstract}



\maketitle

\section{Introduction}

The observed accelerated expansion of the late-time universe, as evidenced by
a host of cosmological data like type Ia supernovae (SN-Ia) \cite{sn1a}, the cosmic
microwave background radiation \cite{cmb} and large scale structure
\cite{lss} came as a great surprise to cosmologists. Although it is straightforward
to explain the effect within the framework of Friedmann-Robertson-Walker cosmology 
by introducing a cosmological constant or a more general (dynamical) dark energy
component, all such explanations give rise to severe coincidence and fine-tuning problems.

An alternative approach postulates that General Relativity is only accurate on
small scales and has to be modified on cosmological distances. This in turn
leads to the observed late-time acceleration of the expansion of the universe
\cite{CD,BDEL,RM,fR}. One of the best-studied examples is the 
Dvali-Gabadadze-Porrati (DGP) brane-world model \cite{dgp}, in 
which the gravity leaks off the 4-dimensional Minkowski brane into the 
5-dimensional bulk Minkowski space-time. On small scales the gravity is bound to 
the 4-dimensional brane 
and the Newtonian gravity is recovered to a good approximation.

One important question is whether such a scenario can be distinguished from one
invoking an invisible dark energy component. It is well known that {\em any}
expansion history (as parametrised by the Hubble parameter $H(t)$) can be
generated by choosing a suitable equation of state for the dark energy
(parameterised by the equation of state parameter $w=p/\rho$ of the dark
energy). This can for example be seen from Eq.~(1) of \cite{bdk}. Let us
illustrate this explicitely for the DGP model, for which the Hubble
parameter evolves as
\be
H^2 -\frac{H}{r_c} = \frac{8\pi G}{3} \rho_m 
\ee
where $r_c$, the crossover scale, separates the 5D and the 4D regimes. It has
to be of the order of $1/H_0$ in order to generate late-time acceleration.
Since matter is conserved on the brane, $\rho_m$ satisfies the
usual conservation equation. Comparing this to the normal Friedmann
equation with an additional dark energy component, we see that we
can move the crossover term to the right hand side and think of
it as a dark energy contribution with $\rho_\de = 3H/(8\pi G r_c)$.
Looking at the conservation equation we find that it is solved
if the effective dark energy has an equation of state given by
\be
1+w_\de = -\frac{\dot{H}}{3H^2} .
\ee
Consequently,
it is impossible to rule out ``dark energy'' based on measurements of
the cosmic expansion history (e.g. SN-Ia data).

Recently there have been claims that it is instead possible to use the
growth rate of structures for this purpose 
\cite{linder1,linder2,polarski,RM} (but see also \cite{linder3} for
cautionary remarks).
This is based on the observation that we can fix the
equation of state parameter $w$ of the dark energy from background data 
and then {\em predict} the evolution of the dark matter perturbations in
a standard cosmological model with dark energy. If the observed growth
rate is different from the predictions, then general relativity with 
dark energy would be ruled out.

However, in this paper we will show that this conclusions makes additional,
very strong assumptions about the nature of the dark energy, and that in general
the growth rate of structure is not sufficient to distinguish between 
dark energy models and modifications of gravity. We will show how the
dark energy perturbations influence the dark matter and the metric
perturbations, and provide an explicit example of a dark
energy model which reproduces the 3+1 dimensional metric perturbations
of the DGP scenario.

\section{Setting the stage}

We start by discussing the fluid perturbations in standard 3+1 dimensional
cosmologies. The perturbations in the energy density are given by
$\delta=\delta\rho/\rho$ and to represent the fluid velocity we use
$V=i k_j T^j_0 /\rho$. Working in the Newtonian (longitudinal) gauge,
the metric can be written as
\be
ds^{2} = -\left( 1+2\psi \right) dt^{2} + a^{2} \left( 1-2\phi\right) dx_{i}dx^{i}
\label{eq:pert_newton_ds}
\ee
with two scalar potentials $\phi$ and $\psi$ describing the perturbations
in the metric. Perturbations in cosmic fluids evolve according to \cite{MaBe}
\bea
\delta' &=& 3(1+w)\phi'-\frac{V}{Ha^2} -\frac{3}{a} 
\left( \frac{\delta p}{\rho} - w \delta \right)\label{eq:delta}\\
V' &=& -(1-3w)\frac{V}{a} + \frac{k^2}{Ha^2} 
\left( \frac{\delta p}{\rho} + (1+w)(\psi-\sigma)\right) \label{eq:V}
\eea
where a prime denotes a derivative with respect to the scale factor $a$.
The physical properties of the fluid are given by the anisotropic stress
$\sigma$ and the pressure perturbation $\delta p$ (in general both can be
functions of $k$). The latter is often
parametrised in terms of the rest-frame sound speed $\cs$,
\be
\delta p = \cs \delta\rho + 3 H a (\cs-c_a^2) \rho \frac{V}{k^2}
\ee
where $c_a^2=\dot{p}/\dot{\rho}$ is the adiabatic sound speed.
Collisionless cold dark 
matter has zero pressure ($w_m=0$), vanishing sound speed ($c_{s,m}^2=0$) and no 
anisotropic stress ($\sigma_m=0$). For the dark energy all these quantities
are a priori unknown functions and have to be measured. For 
the special case of dark energy due to a minimally
coupled scalar field we have a variable $w$ (corresponding to the choice
of the scalar field potential, and fixed by the expansion history of the
universe), $c_{s,\de}^2=1$ and $\sigma=0$ (see e.g.~\cite{hu_lect,KS}).

The perturbations in different fluids are linked via the perturbations
in the metric $\phi$ and $\psi$. Introducing the comoving density perturbation
$\Delta \equiv \delta + 3 H a V/k^2$, their evolution in the standard cosmology
is given by
\bea
k^2\phi &=& -4\pi Ga^2 \sum_{i} \rho_i \Delta_i \label{eq:phi} \\
k^2\left( \phi -\psi\right) &=& 12\pi G a^2\sum_{i}\left(1+w_{i}\right)
\rho_{i}\sigma_{i} \label{eq:psi-phi}
\eea
where the sum runs over matter and dark energy in our case.

The quantity of interest to us is the growth factor $g\equiv \Delta_m/a$
which parameterises the growth of structure in the dark matter. The growth
factor is normalised so that $g=1$ for $a\ll1$ (using that $\Delta_m \propto a$
during matter domination and on sub-horizon scales). For definiteness we
fix $k=200/H_0$ for the numerical results. We assume that $g$ is an observable
quantity (even though of course large scale structure surveys observe
luminous baryonic matter, not dark matter, adding yet another layer of 
complications).

\section{The importance of dark energy perturbations}

\begin{figure}
\epsfig{figure=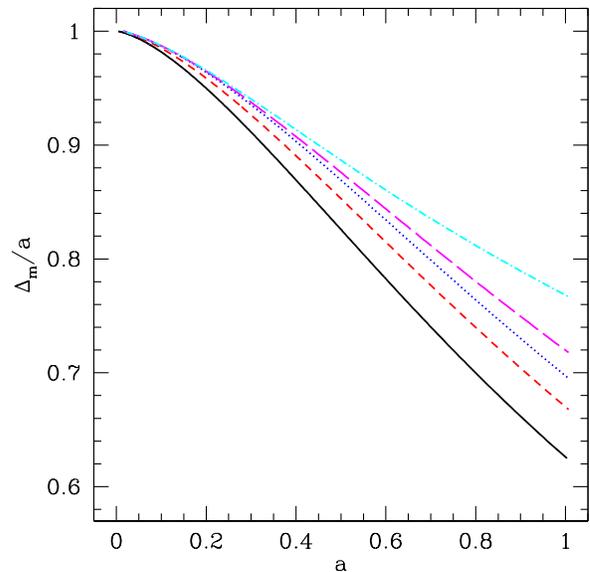,width=3.2in}
\caption{This figure shows how the growth of the matter perturbations depends
on the clustering properties of the dark energy. From the top downward the
sound speed is $\cs=-2\times10^{-4}$ (cyan dash-dotted line),  $\cs=-10^{-4}$ 
(magenta long dashed line), $\cs=0$ (blue dotted line) and $\cs=1$ (red dashed line).
For comparison we also plot the growth factor of the DGP model (black solid line).}
\label{fig:depert}
\end{figure}

We start by noticing that the growth factor is {\em not} uniquely determined
by the expansion history of the universe (and hence $w_\de$). Although the
main effect of the dark energy is to change $H$, leading to $g<1$ at late
times, there is an additional link through the gravitational potential $\psi$.
Different dark energy perturbations will lead to a different
evolution of $\psi$, which can modify the behaviour of $g$. Conventionally
one assumes that the dark energy perturbations are unimportant, e.g. \cite{g2}.
This is a good assumption for scalar field dark energy where the high sound
speed prevents clustering on basically all scales. However, a small sound
speed $c_{s,\de}^2\approx0$ is not excluded. Indeed, it could even be negative, leading
to very rapid growth of the dark energy perturbations. It could also vary in
time. We show in Fig.~\ref{fig:depert} how the growth factor of the dark matter
changes in response to large dark energy perturbations \footnote{We emphasize that
we discuss only how the dark energy perturbations can modify the behaviour
of the dark matter, without taking into account limits from observations.}.

What happens is that, as we decrease the sound speed, the dark energy is able
to cluster more and more. The increased dark energy perturbations lead to enhanced
metric perturbations. The dark matter in turn falls into the potential wells
created by the dark energy, leading to an increase of the growth factor.
Although clearly $g$ is not uniquely determined by $w_\de$, we notice that
it always {\em increases} as we decrease $c_{s,\de}^2$ (at least as long as
the linearised theory is applicable, see also \cite{de_voids}).
Looking at the evolution equations
(\ref{eq:delta}) and (\ref{eq:V}) for $\sigma=0$ ($\Leftrightarrow \phi=\psi$)
we see that the response of the fluids to the metric perturbations is governed by 
the sign of $1+w$. Non-phantom dark energy (as required to mimic the DGP expansion 
history) clusters therefore in fundamentally the same way as the dark matter and 
can only {\em increase} the growth of matter relative to
the case of negligible dark energy perturbations (excluding highly fine-tuned
initial conditions). 

So although the dark energy perturbations can influence the growth factor of
the dark matter, they only seem capable of enhancing it. But Fig.~\ref{fig:depert}
also shows the prediction for the growth factor in the DGP model from \cite{KM},
and it is {\em smaller} than the one of a smooth dark energy component. We therefore
need to change something else if we want to mimic DGP with dark energy. For this
we need to take a closer look at the DGP model.

\section{Anisotropic stress and modified gravity models}

An important aspect of DGP and other brane-world models
is that the dark matter does not see the higher-dimensional
aspects of the theory
as it is bound to the three-dimensional brane. Its evolution is then
{\em the same as in the standard model}. The modifications appear only
in the gravitational sector, represented by the metric perturbations.

The metric perturbation in DGP can be written as \cite{KM,KK}
\bea
k^2 \phi &=& -4\pi G\,a^2\left(1-\frac{1}{3\beta}\right)
\rho_m \Delta_m  \label{eq:dgp_phi}\\
k^2 \psi &=& -4\pi G\,a^2\left(1+\frac{1}{3\beta}\right)
\rho_m \Delta_m \label{eq:dgp_psi}
\eea
where the parameter $\beta$ is defined as:
\be
\beta=1-2r_{c}H\left(1+\frac{\dot{H}}{3H^2}\right)
     = 1+2 r_c H w_\de
\ee

The dark matter does not care if the metric perturbations
are generated (in addition to its own contribution) by
a modification of gravity or by an additional dark energy fluid. Its response
to them is identical. Or to put it differently, if the dark energy and dark
matter together can create the $\phi$ and $\psi$ of
Eqs.~(\ref{eq:dgp_phi}) and (\ref{eq:dgp_psi}) then the growth factor (and
indeed all other cosmological observables) will be the same as in the 
DGP scenario.

\begin{figure}{}
\epsfig{figure=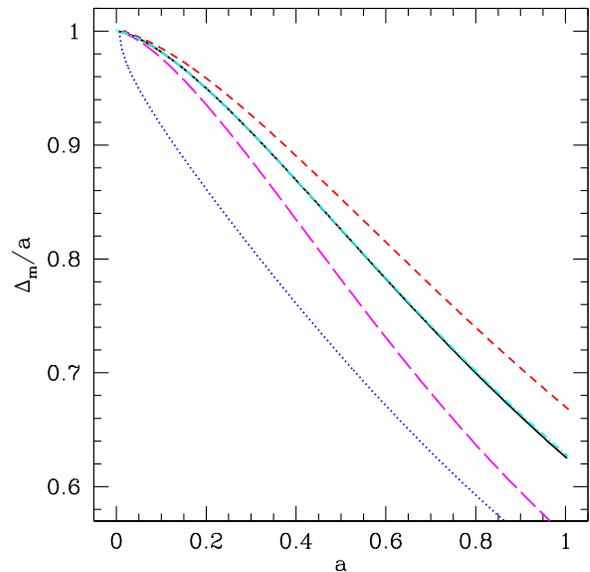,width=3.2in}
\caption{In this figure we show how the anisotropic stress of the
dark energy affects the growth of the dark matter perturbations.
The red dashed line corresponds to scalar field dark energy with
$\cs=1$ and $\sigma=0$. The dotted blue line shows how the dark matter
growth factor decreases for a constant $\sigma_\de=-0.1$. The
long-dashed magenta line uses the theoretical anisotropic stress
of Eq.~(\ref{eq:sigma_theo}) with $\cs=1$, which suppresses the growth
of the matter perturbations too much. Finally, the dash-dotted cyan
line (nearly on top of black solid DGP line) uses the same $\sigma_\de$
but sets the pressure perturbation of the dark energy to 
$\delta p = (1+w)\rho \sigma$ in its rest frame.}
\label{fig:aniso}
\end{figure}

We see immediately that in order to generate these metric perturbations
we will need to introduce an anisotropic stress since $\phi\neq\psi$. This
seems to be a very generic property of modified gravity  that
is also present in $f(R)$ models \cite{fR_aniso} and has been noticed
before. We plot in Fig.~\ref{fig:aniso} again the growth factor for
scalar field dark energy and the DGP model, but now also a family of
dark energy models with non-vanishing anisotropic stress $\sigma$. We
notice that these models can easily suppress the growth of perturbations
in the dark matter for $\sigma<0$ and mimic the behaviour of the DGP model.

Formally we can recover the DGP metric perturbations by choosing 
\be
\sigma_\de = \frac{2}{9\beta (1+w_\de)} \frac{\rho_m}{\rho_\de}
\Delta_m.
\label{eq:sigma_theo}
\ee
for the anisotropic stress of the dark energy, if we can also generate dark
energy perturbations with
\be
\rho_\de \Delta_\de
= -\frac{1}{3\beta} \rho_m  \Delta_m .
\label{eq:de_theo}
\ee
We notice that these are very large dark energy perturbations.
Indeed, if we keep $\cs=1$ and set $\sigma$ to the expression
(\ref{eq:sigma_theo}) we suppress the growth of the matter perturbations
too much, see Fig.~\ref{fig:aniso}. Since $\beta<0$
the large dark energy perturbations of Eq.~(\ref{eq:de_theo})
then increase the matter clustering back to the DGP value.

The required size of the dark energy perturbations in itself is no 
problem, as we can lower the sound speed and even make it negative. 
However, while for $\sigma=0$ we were not able to {\em decrease}
$\Delta_m$ with the help of the dark energy perturbations, 
we find that with a large, negative anisotropic stress we
are unable to {\em increase} it. The required anisotropic stress is
far larger than the gravitational potential $\psi$,
and it starts to be the dominant
source of dark energy clustering in Eq.~(\ref{eq:V}). As it
enters with the opposite sign it now leads to anti-clustering
of the dark energy with respect to the dark matter which
feels only $\psi$ (ie. dark matter overdensities are dark 
energy voids). There is still enough freedom in the choice
of $\sigma$ to match the growth factor very precisely, but
if we could measure both $\phi$ and $\psi$ separately then
we could detect the differences between the two models.

Is it really not possible to match {\em both} $\psi$ and $\phi$
of the DGP model within a generalised fluid dark energy model?
Yes, it is: The metric perturbations have two degrees of freedom, and we do 
have two degrees of freedom of the dark energy to adjust, 
$\sigma$ and $\delta p$. As it turns
out, the parametrisation in terms of the rest-frame sound speed
is too restrictive. This can happen for example if the dark
energy is not composed of a single fluid, see e.g.~the discussion
in \cite{KS}. Allowing free use of the pressure perturbations,
we can choose them for example to cancel the direct effect of
$\sigma$ onto the dark energy perturbation in Eq.~(\ref{eq:V})
by setting $\delta p = (1+w) \rho \sigma$. This reverses the
sign of $\Delta_\de$, and minor adjustments to the pressure
perturbations can then provide the required match to $\Delta_m$.
For the cyan dash-dotted curve in Fig.~\ref{fig:aniso} we
set $\delta p = (1+w) \rho \sigma+3 H a c_a^2 \rho V/k^2$, ie.
we cancelled the contribution of the anisotropic stress in the
dark energy rest frame. This provides a very good solution to
Eqs.~(\ref{eq:sigma_theo}) and (\ref{eq:de_theo}) during matter
domination. It is easy to improve the solution to the point
where it is impossible to distinguish observationally between
the DGP scenario and this generalised dark energy model.

Is linear perturbation theory still valid with such a large
anisotropic stress? Using Eq.~(\ref{eq:de_theo}) we can rewrite
Eq.~(\ref{eq:sigma_theo}) as
$\sigma_\de = -2/(3(1+w_\de))\Delta_\de$.
The anisotropic stress is therefore comparable in size to
$\Delta_m$ and $\Delta_\de$, and at high redshift DGP approaches GR.
It is thus safe to study the dark energy with
linear perturbation theory as long as the dark matter perturbations
stay in the linear regime, even in the presence of the anisotropic
stresses.

\section{Conclusions}

We have shown in this {\em letter} that the growth factor is not
sufficient to distinguish between modified gravity and generalised
dark energy, even if the expansion history (and so the effective
equation of state of the dark energy) has been fixed by observations.
We have also demonstrated that in some cases (notably DGP) the dark
energy can match the metric perturbations completely so that 
cosmological observations cannot distinguish between the two
possibilities.

Although the construction of a matching dark energy model for the
DGP case may seem very fine tuned, we are here more concerned with
the question to what degree this is possible at all. Just measuring
a growth factor that does not agree with scalar field dark energy
is not sufficient to rule out ``dark energy'' and General Relativity. 
But clearly, if
the expansion history {\em and} the growth of matter perturbations
were to match those predicted from a physically motivated and 
self-consistent modified gravity model, a statistical analysis would
rule out a fine tuned dark energy model. However, we should not
forget that as observations seem to indicate $w_\de\approx -1$ it
is rather the modified gravity models that are about to be ruled
out \cite{MM} or look increasingly fine tuned.

\begin{acknowledgments}
M.K. and D.S. are supported by the Swiss NSF. It is a pleasure to
thank Chiara Caprini and Ruth Durrer for interesting discussions,
and Eric Linder and Roy Maartens for comments on the draft.
\end{acknowledgments}

\end{document}